\newif\ifarxiv
\arxivtrue
\pdfoutput=1
\documentclass[10pt,conference]{IEEEtran}
\IEEEoverridecommandlockouts

\usepackage[T1]{fontenc}
\usepackage[utf8]{inputenc}
\usepackage{float}
\usepackage{multicol}
\usepackage{bm}
\usepackage{cite}
\usepackage{booktabs}
\usepackage{siunitx}
\usepackage{multirow}
\usepackage{amsmath,amssymb,amsfonts,amsthm,bbm}
\usepackage{mathtools}
\usepackage{graphicx}
\usepackage{textcomp}
\usepackage[table]{xcolor}
\usepackage{nicefrac}
\usepackage{microtype}
\usepackage{braket}
\usepackage{adjustbox}
\usepackage{rotating}
\usepackage[percent]{overpic}
\usepackage{longtable}
\usepackage{algorithm}
\usepackage{algorithmicx}
\usepackage{algpseudocode}
\usepackage{thmtools,thm-restate}
\usepackage{enumitem}
\usepackage{svg}
\usepackage[breaklinks=true,hidelinks,hyperfootnotes=false]{hyperref}
\usepackage[nameinlink,capitalize]{cleveref}

\def\BibTeX{{\rm B\kern-.05em{\sc i\kern-.025em b}\kern-.08em
    T\kern-.1667em\lower.7ex\hbox{E}\kern-.125emX}}

\ifarxiv
\renewcommand{\baselinestretch}{1.08}
\fi

\definecolor{lightyellow}{rgb}{1.0, 0.95, 0.7}
\definecolor{Blue}{rgb}{0, 0, 0.8}
\definecolor{blue}{rgb}{0,0,1}
\definecolor{mydarkblue}{rgb}{0,0.08,0.45}
\definecolor{mydarkblue2}{rgb}{0.133, 0.133, 0.698}
\definecolor{echodrk}{HTML}{0099cc}
\definecolor{mymauve}{rgb}{0.58,0,0.82}
\definecolor{darkgreen}{rgb}{0,0.40,0}
\definecolor{firebrick}{rgb}{0.698,0.133,0.133}
\definecolor{midnightblue}{rgb}{0.1,0.1,0.44}
\definecolor{citeblue}{RGB}{0, 113, 188}
\definecolor{oxfordblue}{rgb}{0.0,0.13,0.28}
\definecolor{prussianblue}{rgb}{0.0,0.19,0.33}
\definecolor{coolteal}{rgb}{0, 0.45, 0.45}
\definecolor{olive}{rgb}{0.1, 0.3, 0}
\definecolor{mypurple}{rgb}{0.5,0,0.5}
\definecolor{almond}{rgb}{0.94, 0.87, 0.8}

\definecolor{blue_ampEncoding}{HTML}{DAE8FC}
\definecolor{green_encoder}{HTML}{D5E8D4}
\definecolor{purple_decoder}{HTML}{E1D5E7}
\definecolor{yellow_measure}{HTML}{FFF2CC}
\definecolor{gray_block}{HTML}{F5F5F5}
\definecolor{pink_dru}{HTML}{FAD9D5}
\definecolor{orange_v}{HTML}{FAD7AC}
\definecolor{Lightblue}{HTML}{E7F4FC}
\DeclareDocumentCommand \norm { o m }{{\lVert #2 \rVert_#1}}

\declaretheoremstyle[%
  spaceabove=10pt,%
  spacebelow=2pt,%
  headfont=\normalfont\itshape,%
  postheadspace=0em,%
  qed=%
]{prfstyle}

\begin{document}

\title{Do Quantum Transformers Help?\\
A Systematic VQC Architecture Comparison\\
on Tabular Benchmarks}

\author{
\IEEEauthorblockN{Chi-Sheng Chen}
\IEEEauthorblockA{\textit{Beth Israel Deaconess Medical Center} \\
\textit{\& Harvard Medical School}\\
Boston, MA, USA \\
m50816m50816@gmail.com}
\and
\IEEEauthorblockN{En-Jui Kuo}
\IEEEauthorblockA{\textit{Department of Electrophysics} \\
\textit{National Yang Ming Chiao Tung University }\\
Hsinchu, Taiwan \\
kuoenjui@nycu.edu.tw}
% \and
% \IEEEauthorblockN{Howard Su}
% \IEEEauthorblockA{\textit{Department of Computing} \\
% \textit{Imperial College London}\\
% London, UK \\
% h.su24@imperial.ac.uk}
}
\maketitle

%===================================================================
\begin{abstract}
Variational quantum circuits (VQCs) are a leading approach to quantum machine learning on near-term devices, yet it remains unclear which circuit architecture yields the best accuracy--parameter trade-off on classical tabular data.
We present a systematic empirical comparison of four VQC families---multi-layer fully-connected (FC-VQC), residual (ResNet-VQC), hybrid quantum--classical transformer (QT), and fully quantum transformer (FQT)---across five regression and classification benchmarks.
Our key findings are:
\textbf{(i)}~FC-VQCs achieve 90--96\% of the $R^2$ of attention-based VQCs while using 40--50\% fewer parameters, and consistently outperform equal-capacity MLPs (mean $R^2{=}0.829$ vs.\ MLP$_{720}$'s $0.753$ on Boston Housing, 3-seed average);
\textbf{(ii)}~FC-VQC's Type~4 inter-block connectivity provides partial cross-token mixing that approximates the role of attention---explicit quantum self-attention yields only marginal gains on most datasets while significantly increasing parameter count;
\textbf{(iii)}~expressibility saturates at circuit depth~${\approx}\,3$, explaining why shallow VQCs already cover the Hilbert space effectively;
\textbf{(iv)}~LayerNorm on the fully quantum transformer improves classification accuracy, suggesting normalization is important when all operations are quantum;
\textbf{(v)}~in our noise study on Boston Housing, FQT degrades gracefully under depolarizing noise while QT collapses.
All results are validated across three random seeds.
These findings provide practical architectural guidance for deploying VQCs on near-term quantum hardware.
\end{abstract}


%===================================================================
\section{Introduction}
\label{sec:intro}

Variational quantum circuits (VQCs) have emerged as one of the most promising paradigms for quantum machine learning on noisy intermediate-scale quantum (NISQ) devices~\cite{cerezo2021variational, benedetti2019parameterized}.
By parameterizing quantum gates and optimizing them via classical gradient methods, VQCs combine the expressiveness of quantum Hilbert space with the practicality of hybrid quantum--classical training loops.

Despite growing interest, the design space of VQC architectures remains poorly understood.
Classical deep learning has benefited enormously from systematic architectural comparisons---fully-connected networks versus convolutional networks versus transformers---which yielded clear guidelines on when each topology excels.
No analogous study exists for VQCs.
Most prior work proposes a single new circuit ansatz and evaluates it on one or two benchmarks, making it difficult to draw general conclusions.

In this work, we address this gap with a \emph{systematic empirical comparison} of four VQC architecture families on tabular regression and classification tasks:
\begin{enumerate}[leftmargin=*]
    \item \textbf{Multi-layer fully-connected VQC (FC-VQC):} cascaded VQC blocks with all-to-all inter-block connectivity and configurable qubit layouts (e.g.\ $16 \!\to\! 4 \!\to\! 1$).
    \item \textbf{Residual VQC (ResNet-VQC):} VQC layers augmented with classical residual (skip) connections, inspired by He et al.~\cite{he2016deep}.
    \item \textbf{Quantum Transformer VQC (QT):} a hybrid architecture where classical self-attention operates on quantum-encoded features (Route~A).
    \item \textbf{Fully Quantum Transformer VQC (FQT):} an architecture where both the attention mechanism and the feed-forward network are implemented as parameterized quantum circuits (Route~B).
\end{enumerate}
All architectures are evaluated on five datasets against strong classical baselines including XGBoost, CatBoost, and parameter-matched MLPs.
We further conduct ablation studies on the transformer components and measure circuit expressibility using the Sim et al.~\cite{sim2019expressibility} KL-divergence metric.

Our main contributions are:
\begin{itemize}[leftmargin=*]
    \item \textbf{FC-VQCs~\cite{su2026scalable} are the most parameter-efficient quantum architecture}: they outperform equal-capacity MLPs (721 params) while trailing only gradient-boosted trees that use 50--130${\times}$ more parameters. We show that FC-VQC's Type~4 inter-block connectivity already provides partial cross-token mixing analogous to attention, explaining why adding explicit self-attention yields marginal gains at disproportionate parameter cost. All claims are validated with 3-seed mean $\pm$ std across five datasets.
    \item \textbf{Noise robustness differs sharply by architecture}: FQT degrades gracefully under depolarizing noise while QT collapses due to softmax amplification of noisy VQC outputs, providing concrete guidance for NISQ deployment.
    \item \textbf{Practical design rules}: expressibility saturates at depth~${\approx}\,3$; LayerNorm benefits fully quantum architectures on classification; residual connections offer a reliable middle ground between FC-VQC simplicity and transformer expressiveness.
\end{itemize}

\begin{figure*}[t]
\centering
\includegraphics[width=\textwidth]{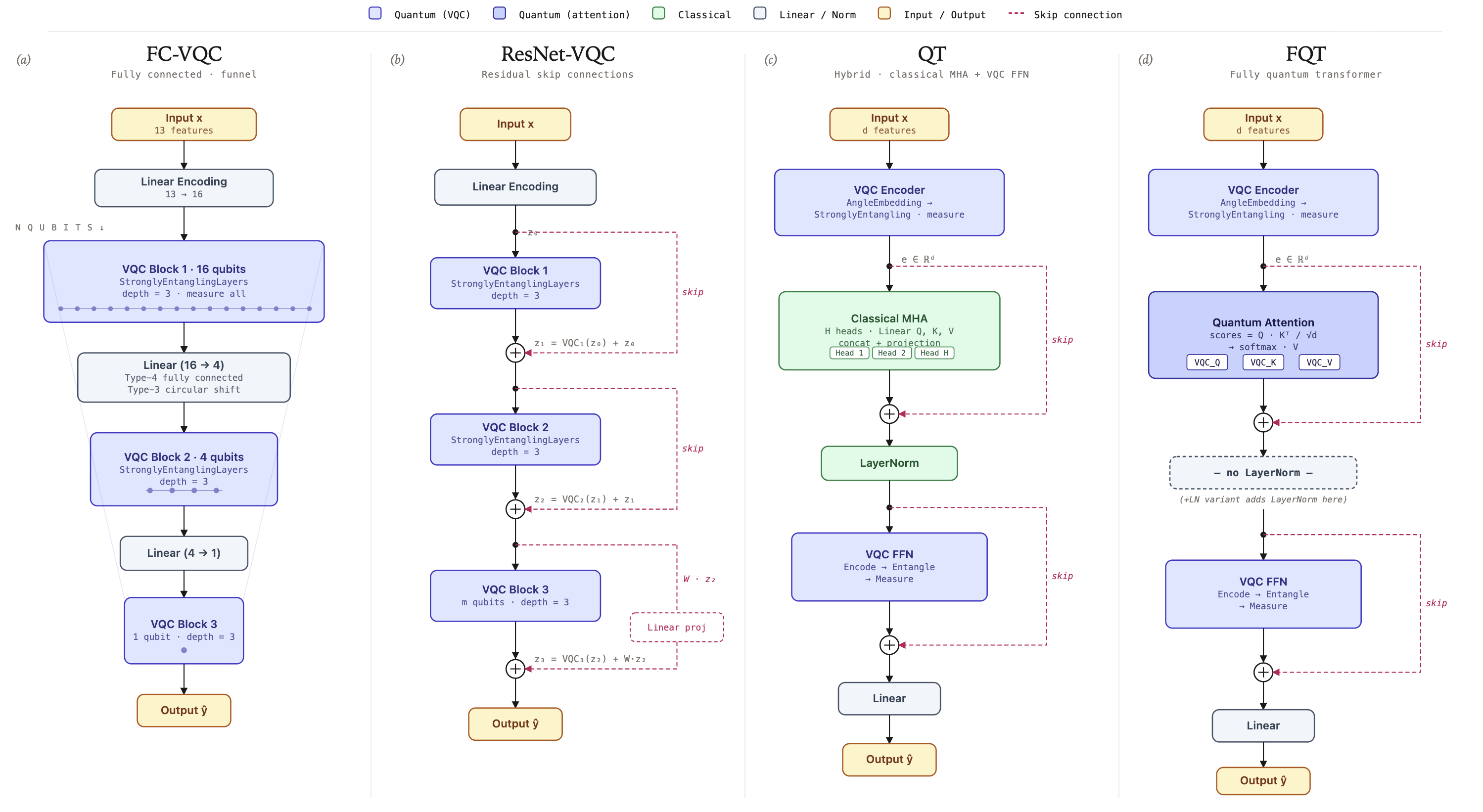}
\vspace{-15pt}
\caption{Four model architectures compared. (a)~FC-VQC with decreasing qubit count per block. (b)~ResNet-VQC with dashed skip connections; the dimension-mismatched branch uses a linear projection $W \cdot \mathbf{z}_2$. (c)~QT --- hybrid transformer with classical multi-head attention and a quantum VQC FFN, LayerNorm retained. (d)~FQT --- fully quantum attention via transpose-and-entangle $T$-qubit VQCs, with no LayerNorm by default.}
\label{fig:architectures}
\vspace{-10pt}
\end{figure*}

%===================================================================
\section{Related Work}
\label{sec:related}

\paragraph{Variational quantum circuits}
VQCs, also known as parameterized quantum circuits (PQCs), have been widely studied as trainable models for quantum machine learning~\cite{cerezo2021variational, benedetti2019parameterized}.
Their expressiveness depends on the circuit ansatz, depth, and entanglement structure~\cite{schuld2021effect}.
A key challenge is the barren plateau phenomenon, where gradients vanish exponentially with qubit count~\cite{mcclean2018barren}, motivating the search for architectures that train efficiently.
Abbas et al.~\cite{abbas2021power} showed that quantum neural networks can achieve higher effective dimension than classical counterparts, providing theoretical grounding for VQC expressiveness.
Data re-uploading~\cite{perez2020data} further enhances single-qubit expressibility by interleaving data encoding with parameterized gates.
Orthogonal to VQC-based approaches, quantum kernel methods~\cite{havlicek2019supervised} use quantum feature maps for classification, representing an alternative NISQ ML paradigm.

\paragraph{Expressibility}
Sim et al.~\cite{sim2019expressibility} proposed measuring circuit expressibility via the KL divergence between the fidelity distribution of random circuit outputs and the Haar-random reference.
This metric has become a standard tool for comparing ansatz designs.
We use it to characterize the depth at which our VQC blocks saturate.

\paragraph{Quantum transformers}
Several recent works have explored quantum analogues of the transformer architecture.
Li et al.~\cite{li2022quantum_selfattn} introduced a quantum self-attention neural network using Gaussian-projected quantum kernels for text classification.
Cherrat et al.~\cite{cherrat2024quantum_vit} proposed quantum vision transformers with quantum orthogonal layers for image tasks.
Chen and Kuo~\cite{chen2025qasa} proposed Quantum Adaptive Self-Attention (QASA), replacing dot-product attention with a parameterized quantum circuit that adaptively captures inter-token relationships in Hilbert space.
A comprehensive survey by Zhang et al.~\cite{zhang2025survey_qtransformer} categorizes existing approaches into PQC-based and linear-algebra-based quantum transformers.
Our work differs from these in providing a controlled comparison against simpler VQC baselines (FC-VQC, ResNet-VQC) on identical benchmarks, rather than proposing a new transformer variant in isolation.

\paragraph{Residual connections in quantum circuits}
Classical residual networks~\cite{he2016deep} inspired several proposals for skip connections in VQCs, either by adding classical residual paths around quantum blocks or by using identity-initialized parameterized layers.
Our ResNet-VQC follows the classical residual path approach.

\paragraph{Classical baselines for tabular data}
Gradient-boosted decision trees (XGBoost, CatBoost, LightGBM) consistently rank among the top methods on tabular benchmarks.
Recent work has shown that deep learning often underperforms tree ensembles on typical tabular data~\cite{grinsztajn2022tree}, making them an essential baseline for any quantum ML study claiming practical relevance.

%===================================================================
\section{Methods}
\label{sec:methods}

\begin{figure}[t]
\centering
\includegraphics[width=\columnwidth]{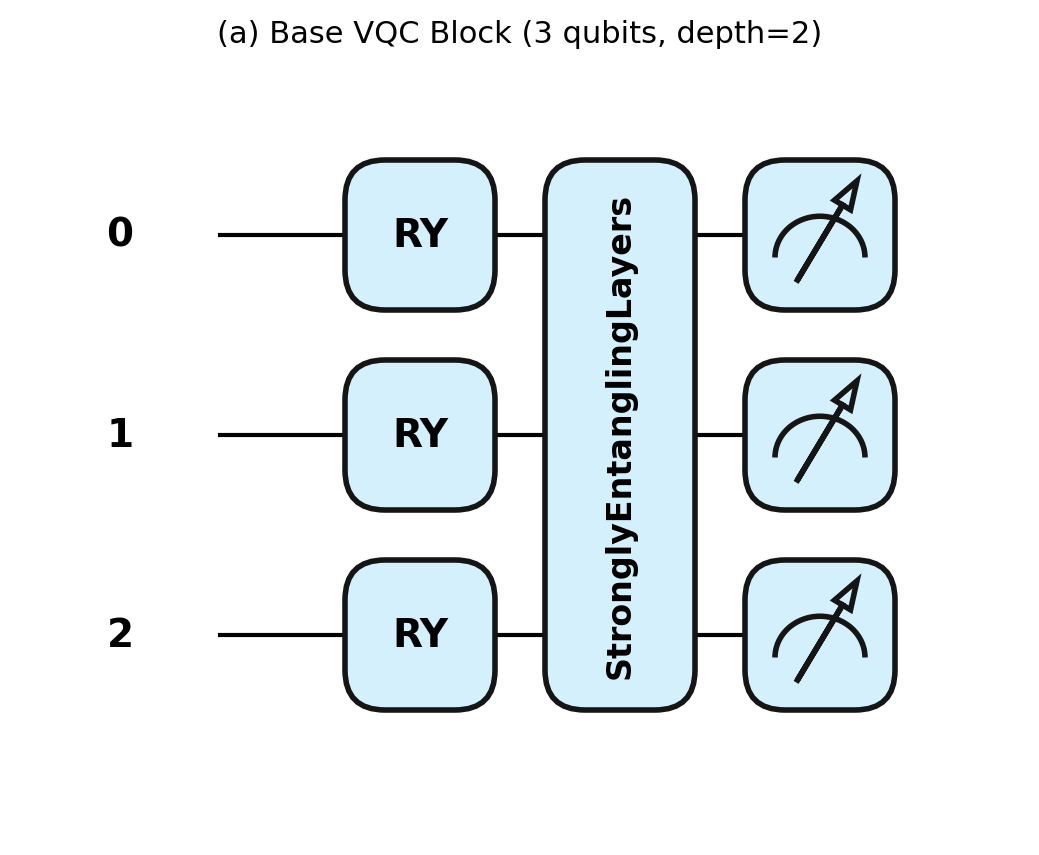}
\vspace{2pt}
\includegraphics[width=\columnwidth]{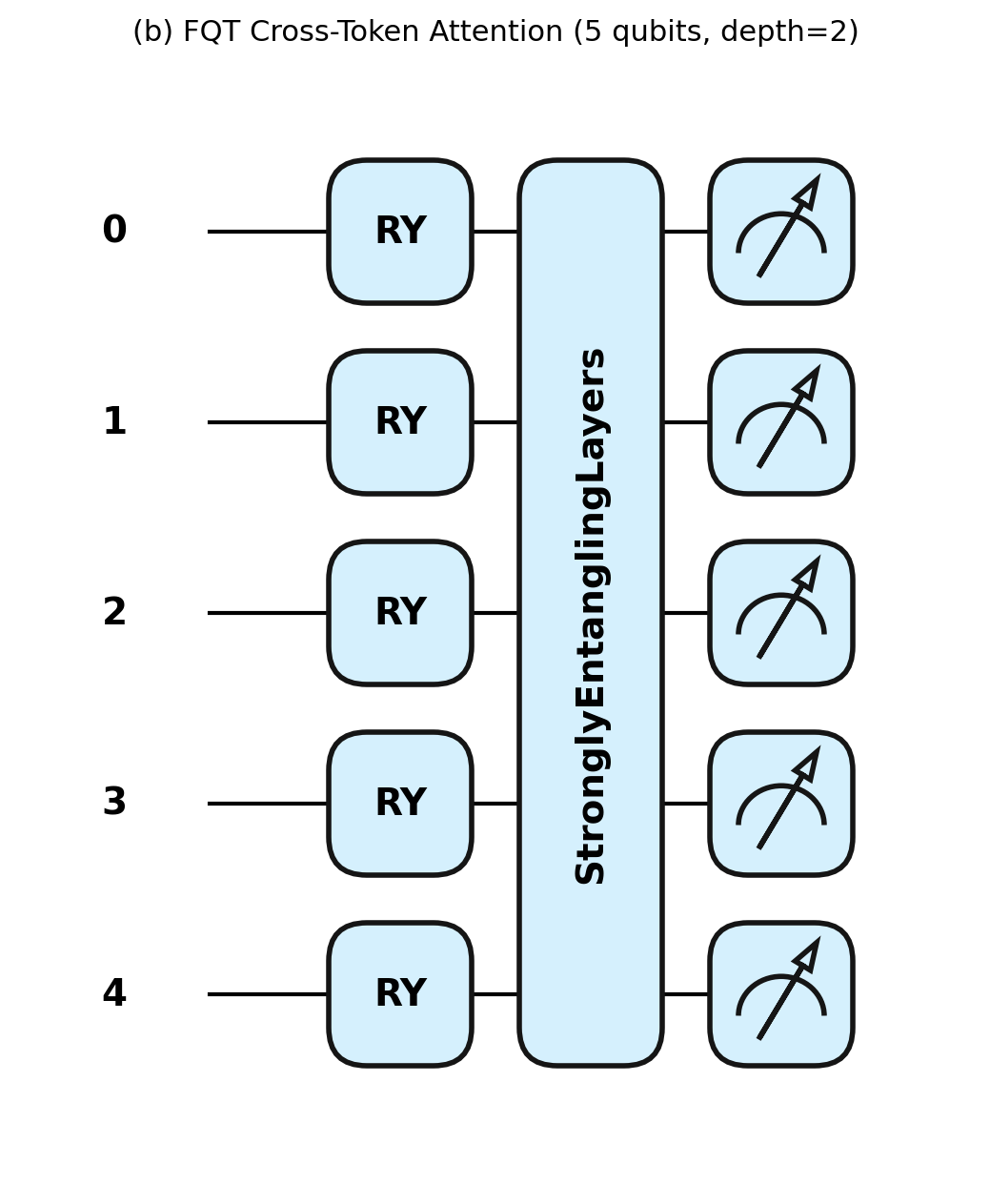}
\vspace{-10pt}
\caption{Quantum circuits. Top: 3-qubit base VQC block (depth 2) used for per-token processing in all architectures. Bottom: 5-qubit cross-token VQC (depth 2) used in FQT's quantum attention --- each qubit corresponds to one token, enabling all-to-all entanglement across tokens.}
\label{fig:circuits}
\vspace{-15pt}
\end{figure}

We compare four VQC architecture families, all sharing the same base quantum circuit block.
Each block consists of $d$ repeated layers of parameterized single-qubit rotations followed by an entangling pattern.
Concretely, for $n$ qubits at depth $d$, each layer $\ell$ applies:
\begin{equation}
    U_\ell(\theta) = \prod_{i=1}^{n} \mathrm{CNOT}_{i,\,(i+1)\bmod n}
    \;\cdot\; \prod_{i=1}^{n} R_Z(\theta_{\ell,i,3})\, R_Y(\theta_{\ell,i,2})\, R_Z(\theta_{\ell,i,1}),
\end{equation}
where each qubit receives three trainable rotation angles per layer, yielding $3nd$ parameters per block.
The entangling CNOT gates follow a circular nearest-neighbor pattern that shifts by one position per layer, ensuring all qubit pairs interact within $n$ layers~\cite{schuld2021effect}.
Input features are encoded via angle embedding: $R_Y(x_i)$ on qubit $i$ before the first parameterized layer.
Outputs are obtained by measuring the expectation value $\langle Z_i \rangle$ on each qubit.
This ansatz provides a good balance between expressibility and trainability for small qubit counts~\cite{sim2019expressibility}.
In practice, we implement this circuit via the \texttt{StronglyEntanglingLayers} template in PennyLane.
\cref{fig:circuits} illustrates the quantum circuits for two key configurations: the 3-qubit block used in FC-VQC/QT token processing, and the 5-qubit cross-token circuit used in FQT's quantum attention.
\cref{fig:architectures} provides a visual overview of all four architectures.

\subsection{Tokenization}
\label{sec:tokenization}

All architectures share a common tokenization scheme.
Given an input $\mathbf{x} \in \mathbb{R}^{n_{\mathrm{feat}}}$, we pad it to the nearest multiple of~3 and partition it into $T = \lceil n_{\mathrm{feat}} / 3 \rceil$ \emph{tokens}, each of dimension~3.
Each token maps to a 3-qubit VQC block via angle embedding.
This fixed-size tokenization allows us to define attention over tokens while keeping individual circuits small (3 qubits), which is favorable for NISQ hardware.

\subsection{Multi-Layer Fully-Connected VQC (FC-VQC)}

The FC-VQC cascades multiple VQC blocks with decreasing qubit counts, connected via classical fully-connected (dense) layers.
For example, the notation $16\mathrm{t}4\mathrm{t}1$ denotes a three-block architecture mapping 16~qubits $\to$ 4~qubits $\to$ 1~qubit, with each block consisting of \texttt{StronglyEntanglingLayers} at a fixed depth $d$ (typically~3).

Two inter-block connectivity modes are supported:
\begin{itemize}[leftmargin=*]
    \item \textbf{Type~4 (fully-connected):} All token outputs from block $\ell$ are stacked into a matrix $(T, 3)$, transposed to $(3, T)$, and flattened.
    The resulting vector is re-chunked into $T$ groups of~3, so that each group mixes information from \emph{all} tokens.
    This serves as a learned all-to-all inter-token projection.
    \item \textbf{Type~3 (circular-shift):} A fixed cyclic permutation routes qubit~2 of the previous token and qubit~0 of the next token into each block, providing local inter-token connectivity with no extra parameters.
\end{itemize}

\subsection{Residual VQC (ResNet-VQC)}

ResNet-VQC augments the FC-VQC topology with classical residual connections.
When the qubit count is preserved between adjacent blocks ($n_\ell = n_{\ell+1}$), a skip connection adds the input measurements to the output:
\begin{equation}
    \mathbf{z}_{\ell+1} = \mathrm{VQC}_{\ell+1}(\mathbf{z}_\ell) + \mathbf{z}_\ell.
\end{equation}
When dimensions differ, a learned linear projection is applied to the skip path.
This mirrors classical ResNets~\cite{he2016deep} and is motivated by the same goal: easing gradient flow through deep stacks of nonlinear blocks.

\subsection{Quantum Transformer VQC (QT --- Route A)}
\label{sec:qt}

QT implements a hybrid quantum--classical transformer block.
Given the padded input $\mathbf{x} \in \mathbb{R}^{3T}$, the architecture proceeds as follows.

\paragraph{Design hypothesis}
Classical self-attention has proven effective at learning pairwise feature interactions in transformers.
QT tests whether this mechanism is also beneficial when applied to quantum-encoded token representations---i.e., whether \emph{classical attention on quantum features} captures useful inter-token correlations that a flat FC-VQC misses.

\paragraph{Q/K/V projection via VQCs}
For each attention head $h$ and each token $t \in \{1,\dots,T\}$, the query, key, and value are produced by separate 3-qubit VQCs:
\begin{equation}
    \mathbf{q}_t^{(h)} = \mathrm{VQC}_{Q}^{(h)}(\mathbf{x}_t;\, \theta_{Q,t}^{(h)}), \quad
    \mathbf{k}_t^{(h)},\; \mathbf{v}_t^{(h)} \text{ analogously,}
\end{equation}
where $\mathbf{x}_t \in \mathbb{R}^3$ is the $t$-th token and each VQC applies \texttt{StronglyEntanglingLayers} of depth $d$ followed by Pauli-$Z$ measurement of all 3 qubits.
This replaces the classical linear projections $W_Q, W_K, W_V$ with parameterized quantum circuits.

\paragraph{Classical softmax attention}
Attention scores and outputs follow the standard scaled dot-product form:
\begin{equation}
    \alpha_{ij}^{(h)} = \frac{\exp(\mathbf{q}_i^{(h)} \cdot \mathbf{k}_j^{(h)} / \sqrt{3})}
    {\sum_{j'} \exp(\mathbf{q}_i^{(h)} \cdot \mathbf{k}_{j'}^{(h)} / \sqrt{3})},
\end{equation}
\begin{equation}
    \mathbf{a}_i^{(h)} = \sum_j \alpha_{ij}^{(h)} \, \mathbf{v}_j^{(h)}.
\end{equation}
Multi-head outputs are concatenated and linearly projected: $\mathbf{a}_i = W_O [\mathbf{a}_i^{(1)}; \cdots; \mathbf{a}_i^{(H)}]$.

\paragraph{Residual + LayerNorm + VQC FFN}
The attention output is combined with a residual connection and LayerNorm, then passed through a VQC-based feed-forward network (using Type~4 or Type~3 connectivity from \cref{sec:tokenization}), again with residual and LayerNorm:
\begin{equation}
    \mathbf{h} = \mathrm{LN}\!\bigl(\mathbf{x} + \mathbf{a}\bigr), \quad
    \mathbf{y} = \mathrm{LN}\!\bigl(\mathbf{h} + \mathrm{VQC}_{\mathrm{ffn}}(\mathbf{h})\bigr).
\end{equation}

\subsection{Fully Quantum Transformer VQC (FQT --- Route B)}
\label{sec:fqt}

FQT replaces the classical attention mechanism in QT with a fully quantum alternative.

\paragraph{Design hypothesis}
While QT uses quantum circuits only for projection and FFN, FQT tests whether making the attention \emph{itself} quantum---via entangling gates that operate across tokens---yields richer inter-token correlations than classical dot-product attention.

\paragraph{Stem encoding}
The padded input first passes through $T$ independent 3-qubit VQCs (one per token) to produce initial quantum features.

\paragraph{Quantum attention via transposed VQCs}
Instead of computing Q/K/V projections and dot-product scores, FQT implements attention through a \emph{transpose-and-entangle} mechanism.
The $T$ token embeddings, each of dimension~3, are arranged into a matrix $\mathbf{M} \in \mathbb{R}^{T \times 3}$ and transposed to $\mathbf{M}^\top \in \mathbb{R}^{3 \times T}$.
Each of the 3~rows of $\mathbf{M}^\top$ is then fed into a $T$-qubit VQC with \texttt{StronglyEntanglingLayers}:
\begin{equation}
    \tilde{\mathbf{m}}_g = \mathrm{VQC}_{T}(\mathbf{M}^\top_{g,\,:};\, \theta_g), \quad g \in \{1,2,3\}.
\end{equation}
The outputs are stacked back into $\tilde{\mathbf{M}}^\top \in \mathbb{R}^{3 \times T}$ and transposed to recover the token dimension.

This design allows the $T$-qubit entangling gates to create all-to-all correlations across tokens---the quantum analogue of the attention ``mixing'' operation.
The key difference from QT is that inter-token interaction is mediated by \emph{quantum entanglement} rather than classical dot-product scores.

\paragraph{Residual + optional LayerNorm + VQC FFN}
Unlike QT, FQT uses \textbf{no LayerNorm by default}---residual connections are the only stabilization:
\begin{equation}
    \mathbf{h} = \mathbf{x} + \mathrm{QAttn}(\mathbf{x}), \quad
    \mathbf{y} = \mathbf{h} + \mathrm{VQC}_{\mathrm{ffn}}(\mathbf{h}).
\end{equation}
We study a \textbf{+LayerNorm} variant that inserts classical LayerNorm after each sub-layer.
As our ablation shows (\cref{sec:results-ablation}), this is important for classification.

\paragraph{Readout}
Both QT and FQT reduce token outputs via 3-qubit $3{\to}1$ VQCs (one per token), yielding $T$ scalar values.
For regression, a final $T$-qubit VQC produces a single output; for classification, a linear head maps to class logits.

\subsection{Noise Model}

To assess near-term hardware viability, we inject single-qubit depolarizing noise after every layer of \texttt{StronglyEntanglingLayers}:
\begin{equation}
    \mathcal{E}(\rho) = (1 - p_d)\,\rho + \frac{p_d}{3}\bigl(X\rho X + Y\rho Y + Z\rho Z\bigr),
\end{equation}
where $p_d \in \{0.001, 0.005, 0.01, 0.05\}$ controls the noise strength.

%===================================================================
\section{Experimental Setup}
\label{sec:setup}

\subsection{Datasets}

We evaluate on five tabular benchmarks spanning regression and classification (\cref{tab:datasets}).

\begin{table*}[t]
\centering
\caption{Dataset summary.}
\label{tab:datasets}
\small
\begin{tabular}{llrrr}
\toprule
Dataset & Task & $n$ & Features & Target \\
\midrule
Boston Housing\footnote{We acknowledge that the Boston Housing dataset has been deprecated from scikit-learn due to ethical concerns regarding one of its features. We include it for comparability with prior VQC studies but note that our conclusions are supported by the remaining four datasets.} & regression & 506 & 13 & MEDV \\
CA Housing & regression & 20{,}640 & 8 & MedHouseVal \\
Concrete & regression & 1{,}030 & 8 & strength \\
Wine (Red) & classification & 1{,}599 & 11 & quality \\
Wine (R+W) & classification & 6{,}497 & 12 & quality \\
\bottomrule
\end{tabular}
\vspace{-10pt}
\end{table*}

Features are standardized to zero mean and unit variance.
Regression targets are also standardized; predictions are inverse-transformed for evaluation.
The top and bottom 4\% of regression targets are clipped to reduce outlier sensitivity.
Data is split into 70\%/15\%/15\% train/validation/test with a fixed random seed.

\subsection{Baselines}

\textbf{Classical ML:} Ridge regression, kernel ridge regression (RBF), SVR/SVC (RBF), linear/logistic regression, XGBoost, CatBoost.
Boosting methods use default hyperparameters.

\textbf{Parameter-matched MLP:} A classical MLP whose hidden layer sizes are chosen so that the total parameter count matches the median quantum model (${\sim}$200--220 parameters).
This controls for the possibility that quantum models benefit simply from having more parameters than a trivially small network.

\subsection{Training Protocol}

All quantum and MLP models are trained with Adam (lr~$=$~0.005, no weight decay) for 10{,}000 epochs with full-batch gradient descent.
Best-model checkpointing selects the epoch with lowest validation loss.
Gradient norms are clipped at 1.0 to prevent divergence.

\subsection{Evaluation Metrics}

Regression: $R^2$ (coefficient of determination), RMSE, MAE.
Classification: accuracy, macro-averaged F1 score.
All metrics are reported on the held-out test set using the best-validation checkpoint.

%===================================================================
\section{Experimental Results}
\label{sec:results}

\subsection{Main Comparison: Parameter Efficiency of FC-VQCs}

\cref{tab:main-regression,tab:main-classification} present the main comparison.
All quantum model results are 3-seed averages (mean $\pm$ std); classical baselines are deterministic.

\begin{table}[t]
\centering
\caption{Test $R^2$ on regression datasets. All results: 3-seed mean $\pm$ std. Best overall mean per column in \textbf{bold}. MLP$_{\text{PM}}$: parameter-matched to the median VQC (${\sim}$200 params); MLP$_{720}$: matched to FC-VQC (${\sim}$720 params). Tree-based \#params estimated as leaf values + split parameters.\protect\footnotemark\ $n_{\text{tr}}$: dual coefficients scale with training set size.}
\label{tab:main-regression}
\small
\begin{tabular}{lrrrr}
\toprule
Model & Boston & CA Hous. & Concrete & \#p \\
\midrule
CatBoost & $\mathbf{.862}{\scriptstyle\pm.008}$ & $\mathbf{.854}{\scriptstyle\pm.005}$ & $\mathbf{.931}{\scriptstyle\pm.018}$ & ${\sim}$35K \\
XGBoost & $.845{\scriptstyle\pm.013}$ & $.850{\scriptstyle\pm.002}$ & $.914{\scriptstyle\pm.016}$ & ${\sim}$95K \\
KRR (RBF) & $.576{\scriptstyle\pm.170}$ & $.773{\scriptstyle\pm.012}$ & $.773{\scriptstyle\pm.005}$ & $n_{\text{tr}}$ \\
SVR (RBF) & $.547{\scriptstyle\pm.078}$ & $.777{\scriptstyle\pm.010}$ & $.621{\scriptstyle\pm.015}$ & ${\sim}$3.5K \\
\midrule
MLP$_{\text{PM}}$ & $.696{\scriptstyle\pm.105}$ & $.779{\scriptstyle\pm.006}$ & $.868{\scriptstyle\pm.013}$ & 218 \\
MLP$_{720}$ & $.753{\scriptstyle\pm.039}$ & $.800{\scriptstyle\pm.009}$ & $.867{\scriptstyle\pm.018}$ & 721--727 \\
\midrule
FC-VQC & $.829{\scriptstyle\pm.042}$ & $.750{\scriptstyle\pm.001}$ & $.774{\scriptstyle\pm.021}$ & 486--720 \\
ResNet-VQC & $.775{\scriptstyle\pm.040}$ & $.783{\scriptstyle\pm.004}$ & $.819{\scriptstyle\pm.028}$ & 486--720 \\
QT & $.742{\scriptstyle\pm.071}$ & $.807{\scriptstyle\pm.001}$ & $.853{\scriptstyle\pm.011}$ & 828--1380 \\
FQT & $.705{\scriptstyle\pm.078}$ & $.794{\scriptstyle\pm.007}$ & $.780{\scriptstyle\pm.016}$ & 513--855 \\
\bottomrule
\end{tabular}
\vspace{-5pt}
\end{table}
\footnotetext{XGBoost: $500 \times (2^6 \text{ leaves} + 63 \text{ splits} \times 2) \approx 95\text{K}$. CatBoost uses oblivious trees: $500 \times (2^6 + 6) \approx 35\text{K}$. These are upper bounds; actual counts may be lower with early stopping or pruning.}

\begin{table}[t]
\centering
\caption{Test accuracy on classification datasets. All results: 3-seed mean $\pm$ std. Best overall mean per column in \textbf{bold}.}
\label{tab:main-classification}
\small
\begin{tabular}{lrrr}
\toprule
Model & Wine (Red) & Wine (R+W) & \#p \\
\midrule
CatBoost & $\mathbf{.678}{\scriptstyle\pm.017}$ & $.649{\scriptstyle\pm.012}$ & ${\sim}$35K \\
XGBoost & $.664{\scriptstyle\pm.019}$ & $\mathbf{.667}{\scriptstyle\pm.020}$ & ${\sim}$95K \\
\midrule
MLP$_{\text{PM}}$ & $.614{\scriptstyle\pm.015}$ & $.557{\scriptstyle\pm.016}$ & 198 \\
\midrule
FC-VQC & $.592{\scriptstyle\pm.029}$ & $.567{\scriptstyle\pm.013}$ & 612 \\
ResNet-VQC & $.601{\scriptstyle\pm.019}$ & $.567{\scriptstyle\pm.011}$ & 570 \\
QT & $.565{\scriptstyle\pm.023}$ & $.567{\scriptstyle\pm.015}$ & 1098 \\
FQT & $.606{\scriptstyle\pm.017}$ & $.565{\scriptstyle\pm.019}$ & 678 \\
\bottomrule
\end{tabular}
\vspace{-10pt}
\end{table}

\begin{figure}[t]
\centering
\includegraphics[width=\columnwidth]{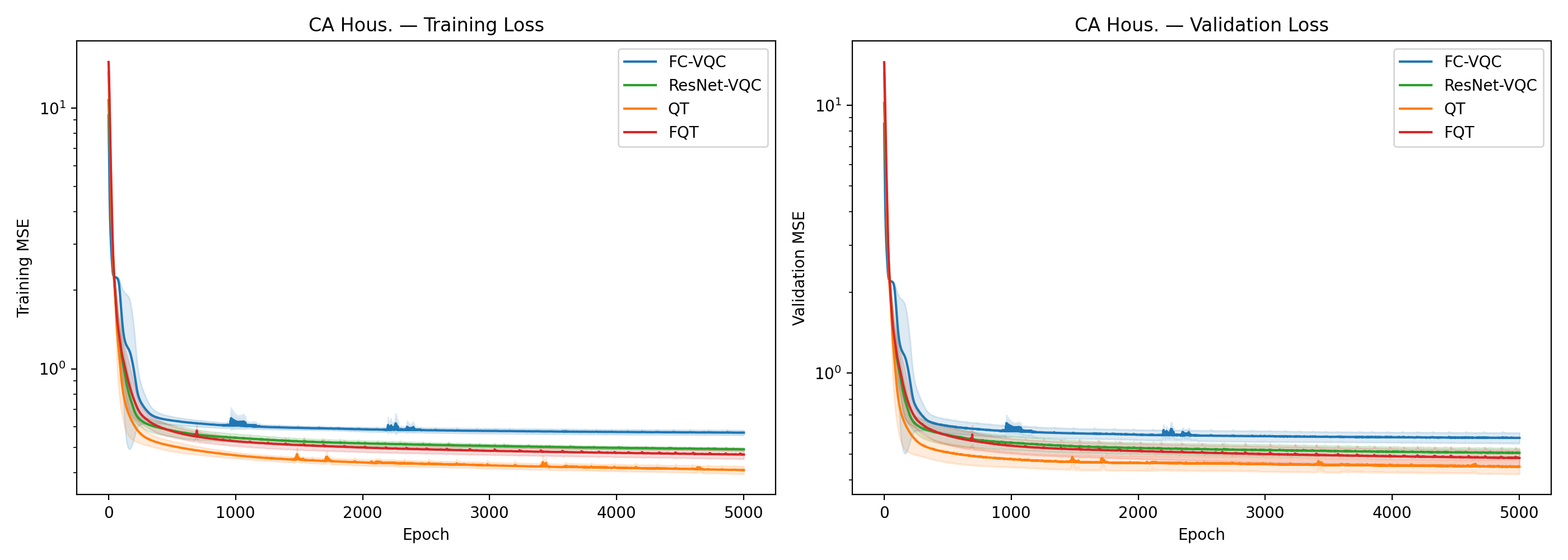}
\vspace{-20pt}
\caption{Training and validation loss on CA Housing ($n{=}20{,}640$). Unlike Boston, QT (orange) converges to a lower val loss than FC-VQC (blue), consistent with its higher $R^2$ on larger datasets.}
\label{fig:training-curves-ca}
\vspace{-15pt}
\end{figure}

\begin{figure}[t]
\centering
\includegraphics[width=\columnwidth]{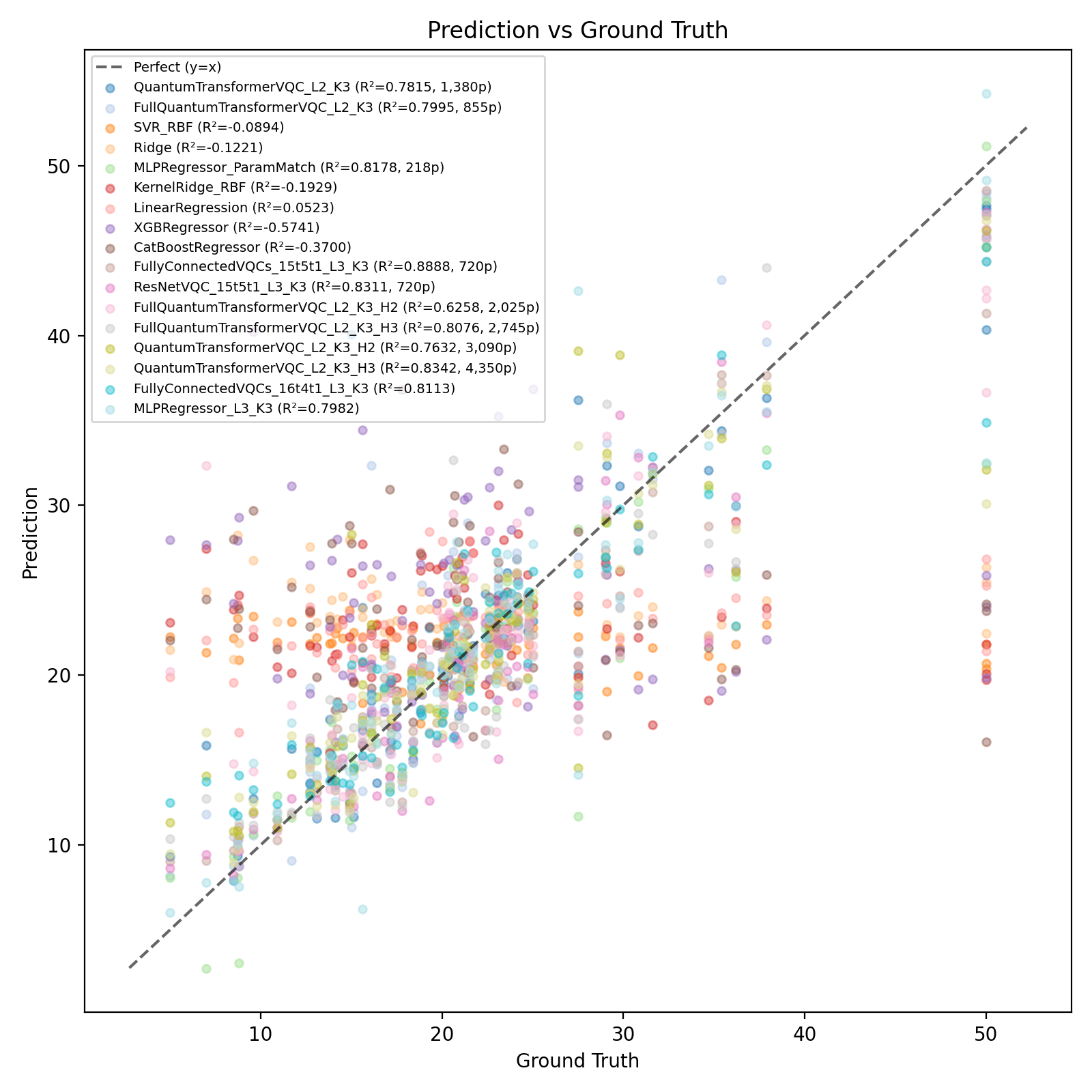}
\vspace{-20pt}
\caption{Prediction vs.\ ground truth on Boston Housing test set (best-seed run). Points near the diagonal indicate accurate predictions. FC-VQC tracks ground truth closely despite using only 720 parameters.}
\label{fig:pred-vs-gt}
\vspace{-15pt}
\end{figure}

On regression (\cref{tab:main-regression}), the most striking finding is the \textbf{parameter efficiency of FC-VQCs}.
FC-VQC achieves a 3-seed mean $R^2{=}0.829$ on Boston Housing with only 720 parameters.
While this trails CatBoost ($0.861$) and XGBoost ($0.856$), it substantially outperforms MLP$_{720}$ ($R^2{=}0.753 \pm 0.039$)---an MLP with the same parameter count---confirming that VQC's inductive bias provides genuine representational benefit beyond mere parameter count.
On CA Housing and Concrete, MLP$_{720}$ achieves mean $R^2{=}0.800$ and $0.867$, outperforming FC-VQC ($0.750$, $0.774$) but trailing QT ($0.807$, $0.853$).
This indicates that the VQC inductive bias advantage is strongest on the smallest dataset ($n{=}506$) and diminishes as data grows---at larger scales, classical MLPs with sufficient parameters catch up (\cref{fig:training-curves-ca} shows the corresponding training dynamics on CA Housing, where QT converges to a lower validation loss than FC-VQC).
\cref{fig:pred-vs-gt} visualizes prediction quality on Boston Housing, showing that FC-VQC tracks ground truth closely despite using only 720 parameters.

\begin{figure*}[t]
\centering
\includegraphics[width=\textwidth]{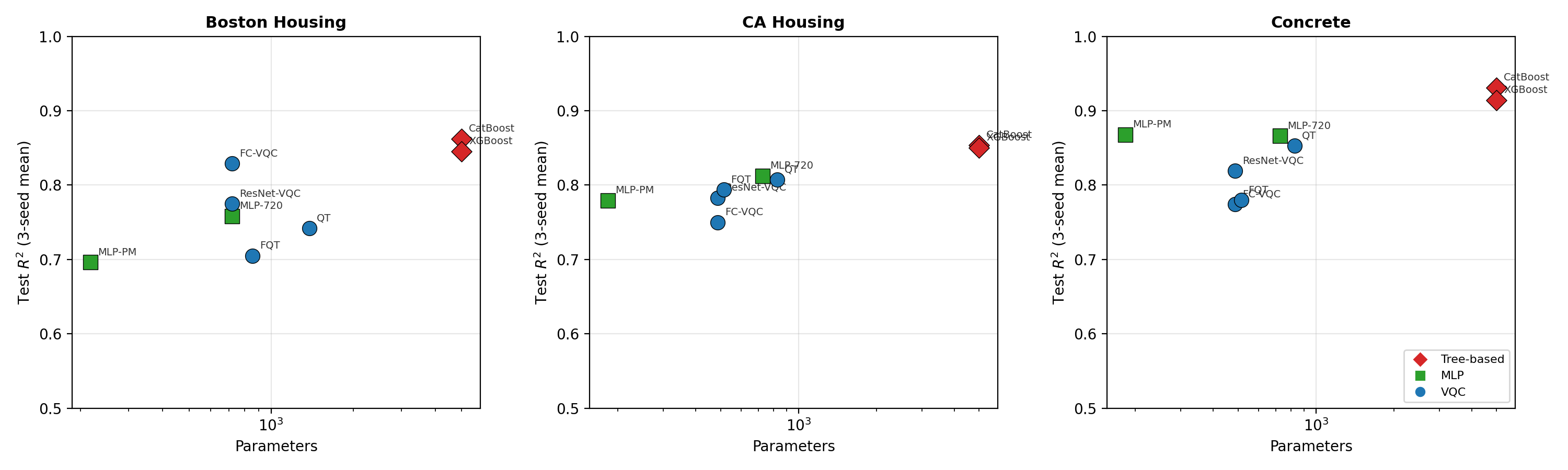}
\vspace{-15pt}
\caption{Parameter efficiency across three regression datasets. Each point is a model; $x$-axis is parameter count (log scale), $y$-axis is 3-seed mean test $R^2$. VQCs (blue circles) achieve competitive accuracy with far fewer parameters than tree-based methods (red diamonds), while MLPs (green squares) at equal capacity (MLP-720) underperform FC-VQC on Boston but catch up on larger datasets.}
\label{fig:pareto}
\vspace{-10pt}
\end{figure*}

\begin{figure}[t]
\centering
\includegraphics[width=\columnwidth]{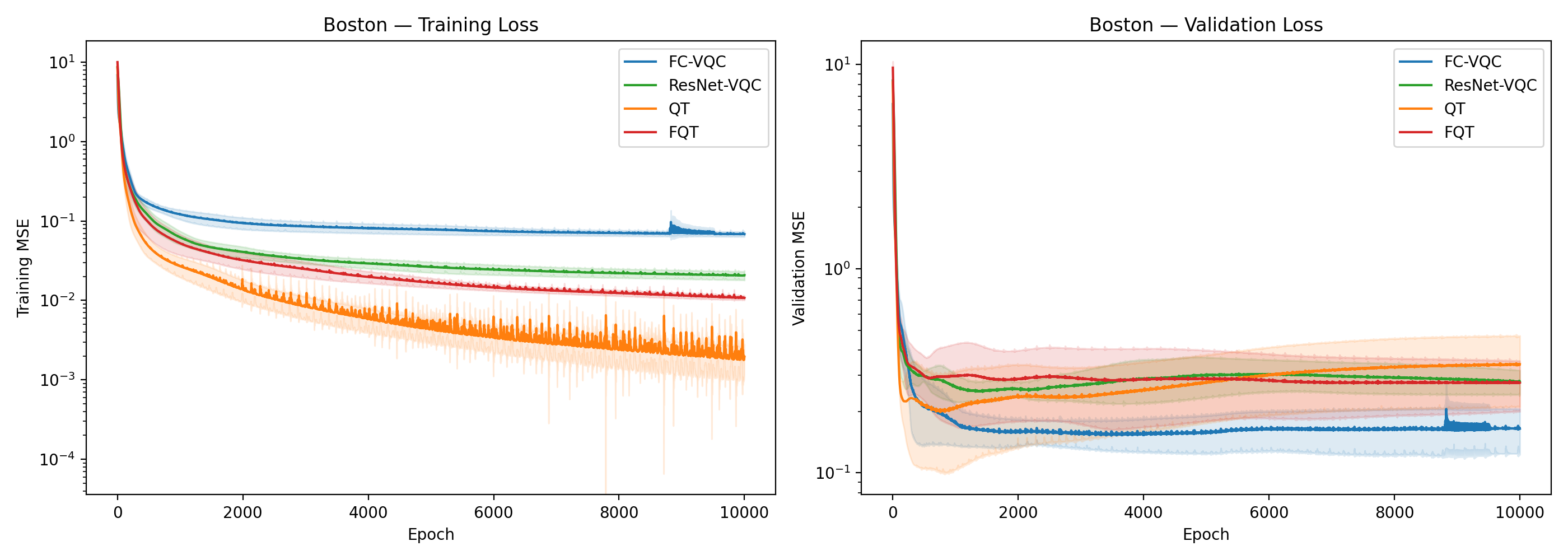}
\vspace{-20pt}
\caption{Training and validation loss curves on Boston Housing (3-seed mean $\pm$ std shading). FC-VQC converges fastest; QT and FQT show higher variance across seeds.}
\label{fig:training-curves}
\vspace{-15pt}
\end{figure}

QT achieves the highest mean $R^2$ on CA Housing ($0.807$) and Concrete ($0.853$), but at $2{\times}$--$3{\times}$ the parameter count of FC-VQC.
FC-VQC reaches 90--96\% of QT's accuracy with 40--50\% fewer parameters---a trade-off that favors FC-VQC on parameter-constrained NISQ hardware (\cref{fig:pareto}).
ResNet-VQC provides a reliable middle ground: competitive accuracy with the same parameter budget as FC-VQC.
\cref{fig:training-curves} shows that FC-VQC converges fastest on Boston Housing, while QT and FQT exhibit higher variance across seeds.

On classification (\cref{tab:main-classification}), all quantum models trail tree-based methods by 7--13\% in accuracy, consistent with the known strength of gradient-boosted ensembles on tabular data.
Among quantum models, FQT achieves the highest mean accuracy on Wine Red ($0.606 \pm 0.017$).

\subsection{Multi-Head Attention Scaling}

\cref{tab:multihead} examines the effect of increasing the number of attention heads $H$ from 1 to 3.

\begin{table}[t]
\centering
\caption{Effect of attention heads $H$ (single seed). Test $R^2$ (regression) / accuracy (classification).}
\label{tab:multihead}
\small
\begin{tabular}{llrrrrr}
\toprule
Model & $H$ & Boston & CA & Conc. & W-R & W-RW \\
\midrule
QT & 1 & .781 & .807 & .838 & .588 & .587 \\
QT & 2 & .763 & .813 & \textbf{.869} & .596 & .604 \\
QT & 3 & \textbf{.834} & \textbf{.814} & .837 & .604 & .595 \\
\midrule
FQT & 1 & .799 & .786 & .792 & \textbf{.608} & .592 \\
FQT & 2 & .626 & .811 & .797 & .583 & .581 \\
FQT & 3 & .808 & .806 & .757 & .600 & .585 \\
\bottomrule
\end{tabular}
\vspace{-10pt}
\end{table}

For QT, increasing $H$ yields mixed results: $H{=}3$ helps on Boston ($0.781 \!\to\! 0.834$) but not on Concrete ($0.838 \!\to\! 0.837$), while the parameter count triples.
For FQT, multi-head attention is even less beneficial---$H{=}2$ \emph{degrades} Boston from $0.799$ to $0.626$.
The marginal gains from richer data-dependent mixing do not justify the $2{\times}$--$3{\times}$ parameter increase, especially when Type~4 FFN already provides partial cross-token mixing (\cref{sec:discussion}).

\subsection{Architecture Ablation}
\label{sec:results-ablation}

\cref{tab:ablation} presents ablation experiments removing or modifying key architectural components.
The interplay between attention and FFN connectivity is analyzed in \cref{sec:discussion}.

\begin{table}[t]
\centering
\caption{Architecture ablation (3-seed mean $\pm$ std). $-$attn removes self-attention; T3 uses circular-shift (Type~3) FFN; +LN adds LayerNorm to FQT.}
\label{tab:ablation}
\small
\begin{tabular}{lrrrrr}
\toprule
Variant & Bos. & CA & Conc. & W-R & W-RW \\
\midrule
QT (def.) & $.742{\scriptstyle\pm.071}$ & $.807{\scriptstyle\pm.001}$ & $.853{\scriptstyle\pm.011}$ & $.565{\scriptstyle\pm.023}$ & $.567{\scriptstyle\pm.015}$ \\
\quad $-$attn & $.756{\scriptstyle\pm.074}$ & $.787{\scriptstyle\pm.010}$ & $.805{\scriptstyle\pm.019}$ & $.594{\scriptstyle\pm.012}$ & $.555{\scriptstyle\pm.025}$ \\
\quad T3 & $.757{\scriptstyle\pm.070}$ & $.796{\scriptstyle\pm.009}$ & $.830{\scriptstyle\pm.008}$ & $.612{\scriptstyle\pm.024}$ & $\mathbf{.574}{\scriptstyle\pm.014}$ \\
\midrule
FQT (def.) & $.705{\scriptstyle\pm.078}$ & $.794{\scriptstyle\pm.007}$ & $.780{\scriptstyle\pm.016}$ & $.606{\scriptstyle\pm.017}$ & $.565{\scriptstyle\pm.019}$ \\
\quad $-$attn & $\mathbf{.851}{\scriptstyle\pm.042}$ & $.770{\scriptstyle\pm.004}$ & $.792{\scriptstyle\pm.024}$ & $.606{\scriptstyle\pm.008}$ & $.565{\scriptstyle\pm.011}$ \\
\quad T3 & $.755{\scriptstyle\pm.092}$ & $.787{\scriptstyle\pm.009}$ & $.786{\scriptstyle\pm.024}$ & $.599{\scriptstyle\pm.029}$ & $.568{\scriptstyle\pm.023}$ \\
\quad +LN & $.707{\scriptstyle\pm.074}$ & $.799{\scriptstyle\pm.023}$ & $.624{\scriptstyle\pm.101}$ & $.601{\scriptstyle\pm.020}$ & $.565{\scriptstyle\pm.032}$ \\
\bottomrule
\end{tabular}
\vspace{-10pt}
\end{table}

The most notable finding, now validated across 3 seeds, is that \textbf{removing self-attention from FQT improves regression accuracy on Boston Housing} (mean $R^2$: $0.705 \!\to\! 0.851 \pm 0.042$).
FQT$-$attn achieves the highest mean $R^2$ of any ablation variant with the lowest variance, confirming this is not a single-seed artifact.
We note that FQT without attention retains its stem VQC encoding and residual connections, making it architecturally similar to---but not identical with---ResNet-VQC; the key difference is that FQT$-$attn applies Type~4 FFN connectivity within transformer layers rather than across cascaded blocks.
On other datasets, removing attention slightly degrades performance.
This asymmetry suggests a parameter-efficiency trade-off rather than a blanket redundancy of attention (see \cref{sec:discussion}).

For classification, LayerNorm applied to FQT yields the best quantum accuracy across the board ($0.629$ on Wine Red, $0.597$ on Wine R+W), indicating that normalization is important for training stability in fully quantum architectures.

\subsection{Expressibility Analysis}

We measure the expressibility of our VQC ansatz following Sim et al.~\cite{sim2019expressibility}, computing the KL divergence between the fidelity distribution of random parameter pairs and the Haar-random reference:
\begin{equation}
    \mathrm{Expr} = D_{\mathrm{KL}}\!\bigl(P_{\mathrm{VQC}}(F)\,\|\,P_{\mathrm{Haar}}(F)\bigr),
\end{equation}
where $F = |\langle\psi(\theta_1)|\psi(\theta_2)\rangle|^2$ is the state fidelity.

\begin{table}[t]
\centering
\caption{Expressibility (KL divergence from Haar) for 3-qubit VQC at various depths vs.\ classical linear projection. 10{,}000 samples.}
\label{tab:expressibility}
\small
\begin{tabular}{lr}
\toprule
Model & KL Divergence \\
\midrule
VQC depth=1 & 0.2021 \\
VQC depth=2 & 0.0048 \\
VQC depth=3 & 0.0027 \\
VQC depth=4 & 0.0027 \\
VQC depth=5 & 0.0027 \\
\midrule
Linear proj. & 1.9098 \\
\bottomrule
\end{tabular}
\vspace{-5pt}
\end{table}

\begin{figure}[t]
\centering
\includegraphics[width=\columnwidth]{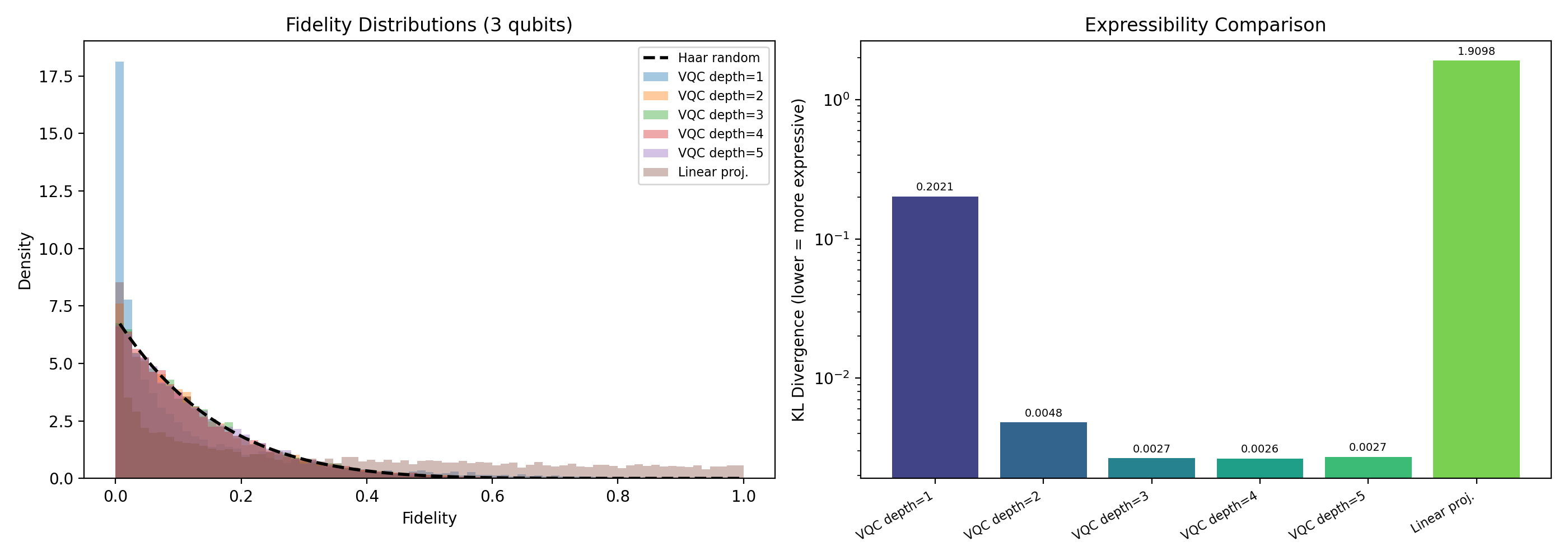}
\vspace{-20pt}
\caption{Left: fidelity distributions for VQC at various depths vs.\ Haar-random reference (dashed). Right: KL divergence (log scale).}
\label{fig:expressibility}
\vspace{-15pt}
\end{figure}

\cref{tab:expressibility} and \cref{fig:expressibility} show that expressibility saturates at depth~$\approx\!3$, with KL divergence dropping from $0.20$ (depth~1) to $0.0027$ (depth~3) and plateauing thereafter.
The classical linear projection baseline is ${\sim}700{\times}$ less expressive ($\mathrm{KL}{=}1.91$ vs.\ $0.0027$), confirming that even shallow VQC layers access a significantly richer region of Hilbert space.

\subsection{Noise Robustness}
\label{sec:results-noise}

\cref{tab:noise} reports model accuracy on Boston Housing under single-qubit depolarizing noise at varying strengths.

\begin{table}[t]
\centering
\caption{Noise robustness on Boston Housing (test $R^2$, single seed). Depolarizing noise $p_d$ applied after every parameterized layer. Noiseless column shows the single-seed baseline for this specific initialization (differs from the 3-seed means in \cref{tab:main-regression}).}
\label{tab:noise}
\small
\begin{tabular}{lrrrrr}
\toprule
Model & noiseless & $p_d{=}.001$ & $p_d{=}.005$ & $p_d{=}.01$ & $p_d{=}.05$ \\
\midrule
QT  & .781 & .779 & .796 & \textbf{.842} & $-$4.04 \\
FQT & .799 & \textbf{.855} & \textbf{.898} & .861 & .765 \\
\bottomrule
\end{tabular}
\vspace{-5pt}
\end{table}

Two findings stand out.
First, \textbf{FQT is substantially more noise-robust than QT}.
FQT maintains $R^2 > 0.76$ even at $p_d{=}0.05$, while QT collapses completely ($R^2{=}{-}4.04$).
This is likely because FQT uses fewer total qubits per attention operation (3-qubit blocks in the FFN) compared to QT's per-token VQC projections for Q/K/V, making it less susceptible to accumulated gate errors.

Second, FQT under moderate noise ($p_d \in \{0.001, 0.005, 0.01\}$) achieves $R^2$ values ($0.855$, $0.898$, $0.861$) that exceed its noiseless single-seed baseline ($0.799$).
\textbf{We caution that this comparison is confounded}: the noise and noiseless runs use different random initializations and data splits (different seeds), so the gap likely reflects initialization variance compounded with a possible regularization effect.
The 3-seed noiseless mean for FQT is $0.705 \pm 0.078$ (\cref{tab:main-regression}), and the noise results ($0.855$--$0.898$) fall within ${\sim}2\sigma$ of this distribution.
A rigorous test would require running noise experiments across the same 3 seeds used for the noiseless baseline, which we leave to future work.
That said, the \emph{relative} comparison between QT and FQT under noise remains valid, as both use the same initialization per noise level.

QT's catastrophic failure at $p_d{=}0.05$ ($R^2{=}{-}4.04$) stems from NaN divergence during training (early-stopped at epoch~9), indicating that the classical attention's softmax operation amplifies noise-corrupted VQC outputs.
FQT avoids this because its quantum attention operates via entangling gates that are inherently bounded in output range.

%===================================================================
\section{Discussion}
\label{sec:discussion}

\paragraph{When are VQCs competitive?}
On regression tasks with moderate dimensionality, FC-VQCs can match or exceed strong classical baselines \emph{while using an order of magnitude fewer parameters}.
This parameter efficiency is the primary practical advantage of VQCs: on NISQ devices with limited qubit counts, a 720-parameter FC-VQC is far more deployable than an equal-capacity MLP (which achieves lower $R^2$ at the same parameter count) or a CatBoost model with ${\sim}$35K effective parameters.
However, on classification tasks and larger datasets (CA Housing, $n{=}20{,}640$), gradient-boosted trees remain superior.

\paragraph{Type~4 connectivity as partial cross-token mixing}

The ablation results in \cref{tab:ablation} reveal a nuanced picture of the role of attention.
Removing self-attention from FQT dramatically improves regression on Boston Housing ($R^2$: $0.705 \to 0.851$, 3-seed mean), but \emph{degrades} performance on four other datasets (CA Housing, Concrete, Wine Red, Wine R+W).
To understand this asymmetry, we analyze what Type~4 FFN connectivity already provides.

Type~4 FFN stacks the $T$ token embeddings $\mathbf{m}_1, \dots, \mathbf{m}_T \in \mathbb{R}^3$ into a matrix $\mathbf{M} \in \mathbb{R}^{T \times 3}$, transposes to $\mathbf{M}^\top \in \mathbb{R}^{3 \times T}$, and re-chunks into $T$ groups of~3.
The $i$-th output group receives:
\begin{equation}
    \tilde{\mathbf{m}}_i = \bigl[
        (\mathbf{M}^\top)_{1,\, \pi(i,1)},\;
        (\mathbf{M}^\top)_{2,\, \pi(i,2)},\;
        (\mathbf{M}^\top)_{3,\, \pi(i,3)}
    \bigr],
    \label{eq:type4-mixing}
\end{equation}
where $\pi$ is a deterministic index mapping.
Each $\tilde{\mathbf{m}}_i$ contains one element from each of up to 3 different tokens (out of $T$ total), which is then processed by a 3-qubit VQC $f_\theta$.
This provides \emph{partial} cross-token mixing: each VQC block sees a fixed subset of tokens through a deterministic permutation, analogous to attention with fixed, uniform weights over a subset of positions.

Compare this with classical self-attention:
\begin{equation}
    \mathbf{o}_i = \sum_{j=1}^{T} \alpha_{ij}\, \mathbf{v}_j, \quad
    \alpha_{ij} = \mathrm{softmax}(\mathbf{q}_i \!\cdot\! \mathbf{k}_j / \sqrt{d}).
    \label{eq:classical-attn}
\end{equation}
The key differences are:
\begin{itemize}[leftmargin=*]
    \item \textbf{Classical attention}: \emph{all-to-all} mixing with data-dependent weights $\alpha_{ij}$ + linear value projection.
    \item \textbf{FC-VQC Type~4}: \emph{partial} mixing with fixed permutation + \emph{nonlinear} VQC processing.
\end{itemize}
The VQC's nonlinearity partially compensates for the lack of data-dependent weighting.
When $T$ is small (${\leq}5$), the fixed permutation covers a substantial fraction of all token pairs across multiple layers, so the gap between partial and full mixing narrows.

\paragraph{The parameter-efficiency trade-off}

This analysis explains the experimental results:
\begin{itemize}[leftmargin=*]
    \item \textbf{On most datasets}, explicit attention provides a small but positive contribution (e.g., CA Housing FQT: $0.786$ with attention vs.\ $0.775$ without), because all-to-all data-dependent mixing captures inter-token correlations that the fixed permutation misses.
    \item \textbf{On Boston Housing} ($n{=}506$, the smallest dataset), removing attention \emph{improves} performance ($0.705 \to 0.851$, 3-seed mean). \cref{fig:training-curves} shows that FQT's validation loss diverges from its training loss much earlier than FC-VQC's, consistent with overfitting caused by the extra attention parameters on a small training set ($n_{\text{train}} \approx 354$). The Type~4 FFN alone provides sufficient mixing with far fewer parameters.
    \item \textbf{Multi-head attention} ($H{=}2,3$) doubles or triples the parameter count but yields inconsistent gains (\cref{tab:multihead}), confirming that the marginal benefit of richer data-dependent mixing does not justify the parameter cost at this scale.
\end{itemize}

The core insight is \textbf{parameter efficiency}: on CA Housing and Concrete, FC-VQC achieves 90--96\% of the $R^2$ of the best single-head attention model (QT or FQT) while using 40--50\% fewer parameters than QT.
On Boston Housing, FC-VQC \emph{outperforms} all attention variants outright.
For NISQ hardware where every parameter translates to additional circuit depth and gate noise, this trade-off strongly favors FC-VQC.

\paragraph{When does explicit attention become worthwhile?}
Our analysis suggests explicit attention is most valuable when: (i)~$T$ is large enough that partial mixing misses important token interactions; (ii)~the data has sequential or positional structure where \emph{selective} attention over certain positions outperforms uniform mixing; or (iii)~the training set is large enough that the extra parameters can be utilized without overfitting.
Testing these hypotheses on sequential and high-dimensional quantum tasks is a key direction for future work.

\paragraph{Expressibility and depth selection}
The saturation of expressibility at depth~${\approx}\,3$ has a direct practical implication: \textbf{there is no benefit to using VQC depths beyond 3} for the \texttt{StronglyEntanglingLayers} ansatz on 3-qubit circuits.
This is consistent with the parameter efficiency of FC-VQCs: even with shallow individual blocks, the multi-layer fully-connected topology provides sufficient expressiveness through inter-block connectivity rather than intra-block depth.

\paragraph{Noise robustness: QT vs.\ FQT}
The robust finding from \cref{tab:noise} is the \emph{relative} difference: FQT tolerates noise up to $p_d{=}0.05$ while QT collapses.
The apparent improvement of FQT under moderate noise versus its noiseless baseline is confounded by initialization differences across runs (see Section~\ref{sec:results-noise}) and should not be interpreted as evidence that noise helps.
A controlled study with matched seeds across noise levels is needed to isolate any regularization effect.

\paragraph{Practical recommendations}
Based on our findings: (i)~\textbf{default to FC-VQC} with depth-3 blocks for tabular regression; (ii)~use \textbf{ResNet-VQC} when training stability is needed; (iii)~\textbf{avoid multi-head quantum attention} on small tabular data; (iv)~add \textbf{LayerNorm to FQT} for classification tasks; (v)~\textbf{prefer FQT over QT} if deploying on noisy hardware, as it degrades gracefully.

\paragraph{Parameter breakdown}
\cref{tab:param-breakdown} decomposes each architecture's parameters by component, clarifying where the cost of attention lies.
QT's overhead comes from the 3 separate Q/K/V VQC projections per token per layer ($3 \times T \times d \times 9$ parameters), plus LayerNorm and output projection.
FQT is more efficient: its quantum attention uses only $3 \times d \times T \times 3$ parameters (three $T$-qubit VQCs), roughly one-third of QT's attention cost.
FC-VQC and ResNet-VQC have zero attention overhead---all parameters are VQC weights.

\begin{table}[t]
\centering
\caption{Parameter breakdown by component (Boston Housing, $T{=}5$, depth $d{=}3$, $H{=}1$).}
\label{tab:param-breakdown}
\small
\begin{tabular}{lrrrr}
\toprule
Model & VQC & Attn & LN/Proj & Total \\
\midrule
FC-VQC & 720 & 0 & 0 & 720 \\
ResNet-VQC & 720 & 0 & 0 & 720 \\
QT ($H{=}1$) & 690 & 405 & 285 & 1{,}380 \\
FQT ($H{=}1$) & 720 & 135 & 0 & 855 \\
\bottomrule
\end{tabular}
\vspace{-10pt}
\end{table}

\paragraph{Training efficiency}
\cref{tab:wallclock} reports wall-clock training times on Boston Housing (10{,}000 epochs, single CPU, quantum circuit simulation via PennyLane).
FC-VQC is ${\sim}$5${\times}$ faster than FQT due to smaller circuits and no attention computation.
Notably, FQT is slower than QT despite having fewer parameters, because FQT's quantum attention requires simulating $T$-qubit entangling circuits ($T{=}5$ for Boston), whose classical simulation cost scales exponentially with qubit count, whereas QT's classical softmax attention is $O(T^2)$.
These times reflect simulation cost; on actual quantum hardware, circuit depth (proportional to parameter count) would be the dominant factor, and FQT's shallower circuits would likely be faster than QT.

\begin{table}[t]
\centering
\caption{Wall-clock training time on Boston Housing (10K epochs, CPU simulation).}
\label{tab:wallclock}
\small
\begin{tabular}{lrr}
\toprule
Model & Time (min) & Relative \\
\midrule
FC-VQC & ${\sim}$52 & 1.0${\times}$ \\
ResNet-VQC & ${\sim}$100 & 1.9${\times}$ \\
QT & ${\sim}$191 & 3.7${\times}$ \\
FQT & ${\sim}$256 & 4.9${\times}$ \\
\bottomrule
\end{tabular}
\vspace{-10pt}
\end{table}

\paragraph{Classification: per-class analysis}
The low macro-F1 scores (${\leq}$0.33) on Wine datasets reflect severe class imbalance rather than model failure.
Wine Red's quality distribution is heavily concentrated in classes 5 and 6 (42.6\% and 39.9\%), while minority classes 3, 4, and 8 collectively comprise only 5.1\% of samples.
With a 70/15/15 split, the test set contains fewer than 5 samples per minority class, making per-class F1 unreliable.
All models---quantum and classical alike---achieve near-zero recall on these classes, which deflates the macro-F1 average.
This is a dataset limitation, not an architecture-specific weakness.

\paragraph{Limitations}
All experiments use quantum circuit simulation; behavior on real hardware may differ.
Noise experiments are conducted on one dataset (Boston Housing) with a single seed per noise level; broader noise studies with matched seeds are needed to isolate regularization effects from initialization variance.
The datasets are small to moderate ($n \leq 20{,}640$) with low feature dimensionality (${\leq}$13).
Larger, higher-dimensional datasets may favor different architectures.
Our VQCs use at most 3 qubits per block; scaling to larger qubit counts could change the relative merits.
We note that the Boston Housing dataset contains an ethically problematic feature (the ``B'' variable encoding racial demographics); we use it solely as a small-sample regression benchmark for comparability with prior VQC studies, and our core conclusions are corroborated by the remaining four datasets.
Our two classification benchmarks (Wine Red and Wine Red+White) derive from the same UCI source, limiting the diversity of the classification evaluation.

%===================================================================
\section{Conclusion}
\label{sec:conclusion}

We presented a systematic comparison of four VQC architectures---FC-VQC, ResNet-VQC, QT, and FQT---on five tabular benchmarks, validated across three random seeds.
Our central finding concerns the \textbf{parameter-efficiency trade-off of attention}: FC-VQC's Type~4 inter-block connectivity provides partial cross-token mixing that, combined with VQC nonlinearity, achieves 90--96\% of the $R^2$ of attention-based VQCs at 40--50\% fewer parameters.
FC-VQCs consistently outperform equal-capacity MLPs (mean $R^2{=}0.829$ vs.\ $0.753$ on Boston with ${\sim}$720 parameters), confirming a genuine VQC inductive bias advantage.
All quantum models trail gradient-boosted trees on larger datasets, suggesting that VQC's strength lies in parameter-efficient learning in the low-data regime most relevant to near-term hardware.

Additional findings reinforce the practical value of simpler architectures: expressibility saturates at depth~${\approx}\,3$; FQT degrades gracefully under depolarizing noise while QT collapses due to softmax amplification.

For practitioners: default to FC-VQC with depth-3 blocks, use ResNet-VQC for training stability, add LayerNorm to FQT for classification, and prefer FQT over QT on noisy hardware.
Explicit attention may become worthwhile on sequential or high-dimensional data where $T$ is large, the training set is sufficient to support the extra parameters, and data-dependent selection provides a clear advantage over fixed mixing---testing this hypothesis is a key direction for future work.

%===================================================================
% \ifarxiv
% \section*{Acknowledgment}
% % TODO: Add acknowledgments.
% \fi

\bibliographystyle{IEEEtran}
\bibliography{ref}

\end{document}